\documentclass[pra,twocolumn,amsmath,amssymb, showpacs]{revtex4}
\usepackage{bm}

\usepackage{epsfig}

\begin{document}

\title{Dimensions, nodes and phases in quantum numbers}

\author{A.\  R.\  P.\  Rau${}^{*}$ \\
Department of Physics and Astronomy, 
Louisiana State University, 
Baton Rouge, Louisiana 70803}

\begin{abstract}

Students of quantum mechanics encounter discrete quantum numbers in a somewhat incoherent and bewildering number of ways. For each physical system studied, quantum numbers seem to be introduced in its own specific way, some enumerating from $1$ and others from $0$, without a common uniting thread. This essay presents a point of view that builds on dimensions, boundary conditions and various inputs that, while known, are often not brought together to present a simple, consistent picture. At the same time, some surprisingly sophisticated connections are also made.
\end{abstract}


\maketitle
    
\section{Introduction}

$\bullet$  Why is there no $1p$ state of the hydrogen atom, the number count beginning only with $2p$? 

$\bullet$  Why are some quantum numbers for bound states counted from $0$ (one-dimensional harmonic oscillator, orbital angular momentum, etc.) while others start at $1$ (particle in a one-dimensional box, Bohr hydrogenic states, etc.)? 

$\bullet$  Why is the two-dimensional hydrogen atom bound more strongly, with four times the binding energy, than the three-dimensional counterpart? It is as if it has a principal quantum number of $n=1/2$ in the Bohr energy formula. 

$\bullet$  Should the Bohr-Sommerfeld quantization condition contain $(n+1/2)$, $(n+3/4)$ or $(n+1)$? When should we use one or the other of these constants, and why?

These, and related questions, occur to almost all students on their first exposure to quantum mechanics. Some are ``trivial", with origins in historical convention and terminology, but even they connect to interesting subtleties of the subject. Others involve surprisingly sophisticated physics. Although the answers to all the above questions are well understood, they lie scattered across books and in the literature. This essay aims to bring them together in a consistent and coherent way which also points to intricacies about the nature of bound states and their connection to scattering states, and the role played by the physical dimensions in which the system resides.

For example, on the first question posed above, Bohr's setting historically the pattern of counting the principal quantum number from 1 predates quantum mechanics (by a dozen years). While this was retained later in atomic physics, the states of the three-dimensional harmonic oscillator are labelled differently; indeed, that system's usage of $1p$, $1d$, etc. sets the terminology in nuclear physics (Section 33 of \cite{ref1}). The difference lies in the convention of using the principal (or energy) quantum number for the atom but the radial quantum number $n_r$ for the isotropic oscillator. The number of radial nodes $n_r$ always starts at zero but the label used for the oscillator is larger by one unit, $n_r+1$, again a matter of convention of no physical significance. But, there is a more subtle reason behind the principal quantum number in the atom being $n_r+\ell+1$ and the unit element here, so that for each $\ell$ it starts one unit larger at $\ell+1$, and therefore at $n=1$ even for $s$-states ($\ell =0$). That extra unit element actually reflects the three dimensions of our world. This is equally true of the isotropic oscillator or the hydrogen atom, depending purely on the physical dimension $D$ of the system. 

More generally, the number is $(D-1)/2$, and it occurs additively to angular momentum, an interplay between dimensions and orbital motion in the kinetic energy that is itself interesting, as will be detailed below. It should be seen as adding to the zero with which otherwise the count would have started for $n$. In turn this accounts for the third question above, this addition being 1/2 for a two-dimensional hydrogen atom, leading to four times the Rydberg unit for the binding energy. Conversely, Coulomb systems in higher dimensions than three have weaker binding, all reflecting this item of a 1/2 unit with each increasing dimension (Section 10.6.1 of \cite{ref2}, and chapter 5, section 3.1 of \cite{ref3}). And, for $D=1$, with $n$ itself really starting at zero, the Bohr expression becomes singular. Indeed, a ``one-dimensional" atom with potential $1/|x|$ would have a pathologically infinite binding (collapse to the origin), although the nature of the singularity is logarithmic and not as in the Bohr expression, arising from the behavior of $\int dx/|x|$ at the origin. Dimensional factors in $r^{D-1}dr$ ``save" the situation for $D >1$, leading to finite binding energy. More details and subtleties of the one-dimensional case lie outside the purview of this essay but have been extensively discussed \cite{ref4,ref5} and applied to atoms in very strong magnetic fields \cite{ref6,ref7}.  

\section{Nodal count and quantum numbers}

For separable problems, the quantum number is directly the count of the number of nodes in the wave function. Therefore, for each separable coordinate, it starts at zero for the nodeless state, proceeding in steps of one along the spectrum as the wave functions acquire successive nodes as required for mutual orthogonality to the wave functions of the states below. The canonical and usually first examples given are the one-dimensional harmonic oscillator or a square potential well. A two-dimensional rotor, with moment of inertia $I$ and azimuthal coordinate $\phi$, has similarly its quantum number, $|m|=0,1, \ldots $, and energy $m^2 \hbar^2/2I $, with $|m|$ the number of nodes in the variable $\phi$. For three dimensional angular momentum, with the additional angular variable $\theta$, each value of $|m|$ is associated with the quantum number $\ell \geq |m|$, $\ell-|m| (=0, 1, \ldots) $ being the number of nodes in $\theta$ of the wave function, the spherical harmonic $Y^{\ell}_m (\theta, \phi)$. Spherical harmonics (Section 28 of \cite{ref1}) have a total of $\ell$ nodes, distributed as $|m|$ in $\phi$ and $\ell -|m|$ in $\theta$. Both of these and their collective sum start at zero and run up through the positive integers.

Turning next to the radial part of problems, as is well known (Section 32 of \cite{ref1}), the kinetic energy of angular motion appears as an effective potential barrier $\ell (\ell +1)\hbar^2/(2mr^2)$ in the radial Schr\"{o}dinger equation. The classical picture of rotational forces pushing a particle away from the origin appears in the quantum wave function as a suppression at small $r$, proportional to $r^{\ell}$. In addition, a new feature enters through the radial part of the Laplacian in the kinetic energy. Both second derivatives in the radial coordinate $r$ and a term in the first derivative occur, the coefficient of the latter, $(D-1)/(2r)$, bringing in the dimensional element. It goes side by side with the factor $r^{D-1}$ that multiplies $dr$ in the volume element in $D$ dimensions. Again, a familiar transformation removes these first derivatives through incorporating a factor $r^{(D-1)/2}$ in the radial wave function, thereby making the radial kinetic energy appear as a single term, $d^2/dr^2$, of a one-dimensional coordinate (albeit with range only from 0 to $\infty$ and not the full line). These matters are usually discussed for the $D=3$ case where this additional factor $r$ adds to the $r^{\ell}$ from angular momentum. This is the origin of the combination $\ell +1$ but the result is generally true for all $D$ giving an additive $(D-1)/2$ to angular momentum $\ell$ (Sections 10.4.2 and 10.7 of \cite{ref2}, and section 10.2 of \cite{ref8}). In dimensions higher than three, this is usually referred to as the ``grand angular momentum" with, as a matter of notation, $\ell$ often replaced by $\Lambda$ and $r$ by the ``hyperspherical " radius $R$ of the higher dimensional space (Section 10.4.2 of \cite{ref2}). As with all angular motion, the number count for $\Lambda$ starts at zero. The radial nodal count also starts at zero but the additive $(D-1)/2$ is the minimum value that appears in the energy expression.

The removal of the first derivatives in radial coordinate comes at a cost, although this is obscured by the accident that in three dimensions, where it is most discussed, this cost seems to be zero, the $(2/r)d/dr$ term simply removed upon multiplying the radial wave function by $r$. In all other dimensions (except, of course, $D=1$ when the question does not arise), this transformation to $R^{(D-1)/2}$ times the radial function is accompanied by the appearance of an additional $(\hbar^2/2m)[(D-1)(D-3)/4R^2]$ ``effective potential" term in the resulting Schr\"{o}dinger equation (Section 10.2 of \cite{ref8}). As just noted, it vanishes of course for $D=1$ but also for $D=3$. In all other dimensions, it appears as an angular momentum-like term with $(D-3)/2$ adding to the usual, or ``hyperspherical grand", angular momentum. Note that it is attractive for $D=2$, repulsive in all other cases and specifically for dimensions higher than three.

This occurrence of a dimensional element additively with angular momentum has many consequences. It has been used for perturbative calculations that start with the large dimension limit which often simplifies a given problem \cite{ref9,ref10,ref11,ref12}. The effective $1/R^2$ potential is the origin of the very interesting ``Efimov effect" \cite{ref13} in which three bodies can bind even though the pairwise attractive potentials between any of the three pairs are insufficient to form two-body bound states. Few-body problems for $N$ particles can be regarded as problems in $3(N-1)$ dimensions involving such a hyperspherical term. The purely kinematical $1/R^2$ potential is independent of the actual physical potentials between the bodies so that the Efimov phenomenon is largely independent of whether one is dealing with short or long range or other details of those potentials and is a universal phenomenon \cite{ref14,ref15}. The behavior of scattering cross sections just above reaction thresholds also depends critically on dimensions because of the phase space involved \cite{ref16}.

\section{Bound states as poles of the scattering matrix}

The same feature discussed in the above section, of a dimensional unit adding to angular momentum, also appears in the more sophisticated and complete quantum-mechanical treatment of the Coulomb problem wherein two bodies interact through the Coulomb potential $1/r$. The Bohr energy levels may be viewed as poles of the Coulomb scattering amplitude in the complex energy plane, the crucial element being a factor $\Gamma (\ell +1 -iZ/ka_0)$ in the denominator, where $(-Ze^2)$ is the product of the charges, $a_0$ the Bohr radius, and $k$ the wave number given by $E=\hbar^2k^2/2m$ (section 136 of \cite{ref1}, section 4 of \cite{ref17}, section 14.6 of \cite{ref18}). For the attractive case, $Z >0$ stands for the nuclear charge seen by the electron in a hydrogenic atom. 

Extrapolating to negative energies with $k =i\kappa$, this gamma function has zeroes for $\kappa =Z/na_0$, so that the scattering amplitude or S-matrix has poles at those positions. These are precisely the Bohr energy levels, $E=-\hbar^2 Z^2/(2ma_0n^2)$ (section 136 of \cite{ref1}, section 7.7 of \cite{ref19}, section 14.6 of \cite{ref18}). Alternatively, with the S-matrix written as a ratio of two ``Jost" matrices (Sections 5.3 and 7.3 of \cite{ref2}), zeroes of the Jost matrix in the denominator mark the bound state energy levels (Section 5.3.2 of \cite{ref2}, and section 14.6 of \cite{ref18}). In all this, the unit element that stands in the gamma function is again the $(D-1)/2$ discussed in the previous section. Thus, the zeroes of the gamma function occur at values of $n$ larger than $\ell$ by one unit, that dimensional element. The result immediately generalizes to other values of $D$, the two-dimensional atom therefore having an effective quantum number of 1/2 and four times the binding whereas for $D >3$, the energies are correspondingly smaller for the ground state (Chapter 5, section 3.1 of \cite{ref3}). Note that we refer to a $D$-dimensional Coulomb problem as one with a potential varying inversely as the radius $R$, not as solutions of Laplace's equation in those dimensions in which case the radial dependences would vary.

This same gamma function is the crucial component of the so-called Coulomb normalization or ``Gamow enhancement" factor (section 136 of \cite{ref1} and section14.6 of \cite{ref18}) that, for the attractive Coulomb problem, enhances wave function density at small $r$ and is manifest in several phenomena (Sections 2.5 and 5.7 of \cite{ref2}). In particular, the derivative $dE/dn$, which places normalization of bound and continuum states on par and thereby allows smooth extrapolation of spectroscopic data into the scattering continuum, is proportional to $1/n^3$ for the Coulomb problem. As a result, radial wave functions of bound states all carry this familiar $1/n^{3/2}$ factor. A whole body of work called quantum defect theory (QDT) systematizes these effects of long range forces such as the Coulomb but more generally (Chapter 5 of \cite{ref2}). An amplitude factor, called $B$ in this QDT, for positive energies and its counterpart $A$ at negative energy similarly involve the gamma function discussed above or its counterparts for other potentials. Jost matrices can be expressed in terms of them and again, bound state positions for other long range potentials at zeroes of those Jost functions systematized (see chapter 5 and Table 5.1 of \cite{ref2}).

\section{Bohr-Sommerfeld quantization, Jeffreys-Wentzel-Kramers-Brillouin (JWKB) method, and subtleties about phases}

For each separable coordinate in any problem, the Bohr-Sommerfeld condition,

\begin{equation}
\oint p(x) dx = (n+c)h,
\label{eqn1}
\end{equation}
provides a simple route to quantization (Section 48 of \cite{ref1} and section 2.4 of \cite{ref19}). The integral is over a closed loop in the classically allowed region between turning points (see Fig. 1), and $c$ a constant, usually taken to be 1/2 in a situation as shown in the figure. This simple formula gives the correct energy levels of the one-dimensional harmonic oscillator all the way down to the zero point energy of the ground state. It is derived usually as an application of the JWKB method and of its ``connection formulae" for bound states, where the connection between JWKB wave functions on either side of the turning point introduces phases of $\pi/4$ each to add to the accumulated phase represented by the integral over the wave number $p/\hbar$ in the classically allowed region of wave oscillations (Section 47 of \cite{ref1}). This makes clear the connection between $n$ and the number of nodes in that oscillation, with $n$ starting at 0. Note that it is the Planck constant $h$, not $\hbar$, that appears in Eq.~(\ref{eqn1}).

\begin{figure}
\scalebox{2.0}{\includegraphics[width=1.6in]{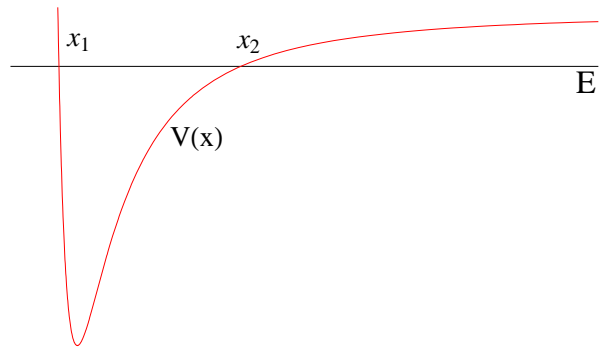}}
\caption{Sketch of a potential well and a bound state at energy $E$. Turning points are indicated. Courtesy: Sai Vinjanmapathy}
\end{figure}

It is also well known that the constant in Eq.~(\ref{eqn1}) is 3/4 when the potential rises abruptly to infinity as at a hard wall, on one or the other side. This is the case for what are called ``half-space" problems, as with an oscillator term for $x >0$ and infinite potential for $x \leq 0$. The requirement that the wave function must vanish at a hard wall retains only the odd parity solutions of the oscillator on the full line, with a spectrum starting at $3\hbar \omega/2$ and going up in steps of $2\hbar \omega$. This is precisely the result given by Eq.~(\ref{eqn1}) with $c=3/4$. The justification for this choice is also clear in the JWKB derivation because the wave function must now vanish at the left turning point at the origin and for all $x \leq 0$ which negates the $\pi/4$ phase contribution from that end. 

Going further, clearly when there is a hard wall at both ends, as in the problem of a particle in a box, both $\pi/4$ contributions drop out and $c =1$ in Eq.~(\ref{eqn1}). Once again, this choice gives the exact energy levels of the particle in a box and also shows that while the nodal quantum number count starts at zero, it is because of this additional unit element that the energy expression seems to start with $n=1$.

The discussion of different values of $c$ might seem pedantic, given that the JWKB method is semi-classical in nature and expected to be valid only for large quantum numbers $n$. But, it is clear from the above examples that there are good reasons for the various choices of $c$, and that they give exact results all the way down to $n=0$. This argues for a more basic reason than the specifics of the JWKB method. Indeed, the basic origin of this $\pi/4$ phase contribution is that it is the fourth root of $(-1)$. The point is that across a turning point, the expression for the local kinetic energy, $E-V(x)$, changes sign. The wave function is proportional to the inverse square root of the local momentum $p(x)$, the quantum-mechanical realization of the classical expectation that the probability of finding a particle at a location is inversely proportional to the velocity, or equivalently, its momentum. This gives rise to the fourth root of $(-1)$ in going from classically allowed to classically forbidden region (see a footnote on p. 107 of \cite{ref19}, section 47 of \cite{ref1}, and section 5.8 of \cite{ref2}). Of course, with hard walls, when the wave function is identically zero on one side and the derivative on the two sides no longer has to be matched, there is no connection involved, merely that the wave function starts at zero at that point. 

This same result is manifest at a much more sophisticated level in quantum defect theory. It follows from analytic properties of the Jost functions that the amplitude and phase factors, $B(k)$ and $\eta(k)$, of QDT, upon extrapolation to the bound state sector of negative energies through $k \rightarrow i\kappa$, satisfy the universal relation valid for any potential \cite{ref20},

\begin{equation}
arg \{B^{1/2}(i\kappa) \exp[2i\eta(i\kappa)]\} =\pi/4.
\label{eqn2}
\end{equation}

There are further levels of sophistication regarding the constant $c$ in Eq.~(\ref{eqn1}) \cite{ref21,ref22}, which is called the ``Maslov index", including what is called the ``uniform approximation" in JWKB theory \cite{ref23,ref24}. These lie outside the scope of this essay which is concerned only with the simple phase considerations underlying the values of 1/2, 3/4 and 1, and their origin in the fourth square root of $(-1)$.              

\section{Summary}

Simple questions about the numbering of quantum numbers and of the nature of binding energy involve features about nodes and phases of wave functions, the dimensions of the system, and analytic properties connecting scattering and bound states. In discussing these, sophisticated connections can be made to fundamental features of quantum physics.  

This essay is dedicated to the memory of a good friend and colleague, Mitio Inokuti. A distinguished radiation physicist and founding correspondent of the Comments on Atomic and Molecular Physics, he put great stress on accuracy and clarity in his research and pedagogical writings. It was his question that opens this essay. He asked it at my last meeting with him a little before his death. Following a discussion that covered some of what is presented here, he encouraged that it be presented for publication. While he did not live to see this essay, it expresses my respect and admiration for him and his physics.

This work has been supported by a Roy P. Daniels Professorship at LSU. I also thank the Humboldt Foundation for its support and Prof. Gernot Alber and the Technical University, Darmstadt, for their hospitality during its writing.

\end{document}